\documentclass[
preprint,
 amsmath,amssymb,
 aps,
floatfix,
]{revtex4-2}
\usepackage{graphicx}
\usepackage{braket}
\begin{document}
\preprint{}
\title{Rydberg Atom Sensors in Multichromatic Radio Frequency Fields}
\author{Mohammad Noaman, Donald W. Booth and James P. Shaffer}

\affiliation{Quantum Valley Ideas Laboratories, 485 Wes Graham Way, Waterloo, ON N2L 6R1, Canada}

\date{\today}

\date{\today}

\begin{abstract}
Rydberg atom-based sensors are a new type of radio frequency sensor that is inherently quantum mechanical. Several configurations of the sensor use a local oscillator to determine the properties of the target radio frequency field. We explain how the physics of Rydberg atom-based sensors in two or more radio frequency fields can be precisely described by a multiply dressed Jaynes-Cummings model. Studying Rydberg atom-based sensors in two or more near resonant radio frequency fields is important for understanding how interfering signals as well as the local oscillator can affect measurements. Studies, so far, focus on a simplified approximation for the local oscillator-target field interaction that uses an analogy to radio frequency heterodyning. The atom acts as a medium for exchanging electromagnetic field excitations of the field modes whose spectrum is a ladder. The Jaynes-Cummings states and their avoided crossings can be used to determine the properties of the radio frequency fields. Radio frequency field sensitivity enhancement for non-resonant radio frequencies is achieved and self-calibrated measurements are recovered under specific conditions described by the theory.
\end{abstract}
\maketitle

Rydberg atom based sensors have seen growing interest for a range of applications in metrology and radio frequency (RF) field sensing~\cite{Kumar:2017, Fan:2015, Meyer:2020, Adams:2020, Bohaichuck:2022, Booth:2022}. Accessibility to transition frequencies from sub-GHz to sub-THz makes these sensors suitable for calibration-free measurements over a wide range of RF frequencies. In a typical experiment, the properties of an unknown RF field, resonant with a pair of Rydberg states, is estimated by observing Autler-Townes (AT) splitting. We study the situation where a second RF field, slightly detuned, addresses the same pair of Rydberg states. In one case the stronger of the two fields acts as a target field while the weaker one acts as an interferer. In a second case, the stronger field can be utilized as a control field while the weaker one acts as a target field. The choice between the two cases depends on the type of measurement that is carried out. Taking the former case, the target field splits the EIT feature into two AT peaks. The interferer produces additional peaks and splittings for various detunings and field strengths. The emergence of the additional spectral features is the result of harmonic and sub-harmonic resonances that can be explained by the doubly dressed Jaynes-Cummings model. The atoms act as a medium for the exchange of RF excitations between the two RF fields.

The understanding of RF dressing through the Jaynes-Cummings model enables recovery of self-calibration for measuring the properties of the RF fields, including the off-resonant interferer. Off-resonant sensing can be achieved by other means, e.g., by tuning the transition frequencies using a DC field, but these approaches have not demonstrated SI-traceability. In our approach, interfering and target signals can be measured with SI-traceability. Several studies of two level atoms starting from the ground state and interacting with two optical fields appear in the literature~\cite{Agarwal:91, Greentree:1999, Gustin:2021, Papageorge:2012, Rudolph:98, Wang:2003}. These studies do not address an understanding of an interfering RF field in Rydberg atom-based sensors. Work on Rydberg atoms in two RF fields for sensing applications has been done~\cite{Jing:2020, Hu:2022, Gordon:2019, Rotunno:2023}, but the focus in these cases is on treating the two fields analogous to a classical RF mixer. Although these approaches are valid for certain specific range of parameters, a general response of the Rydberg sensor to a bichromatic field is more complicated.

To understand Rydberg sensors in a bichromatic RF field, we carry out experiments and compare them to numerical calculations of the Rydberg EIT system coupled to two RF fields. The experimental data, analytical expressions and numerical calculations demonstrate a calibration-free RF sensing of the target and perturbing field. An understanding of the results is obtained by adopting a multiply dressed Jaynes-Cummings picture. Our studies are not limited to two fields, but can be generalized to multiple RF fields via a multi-mode Jaynes-Cummings model. 

\begin{figure}
    \includegraphics[scale=1]{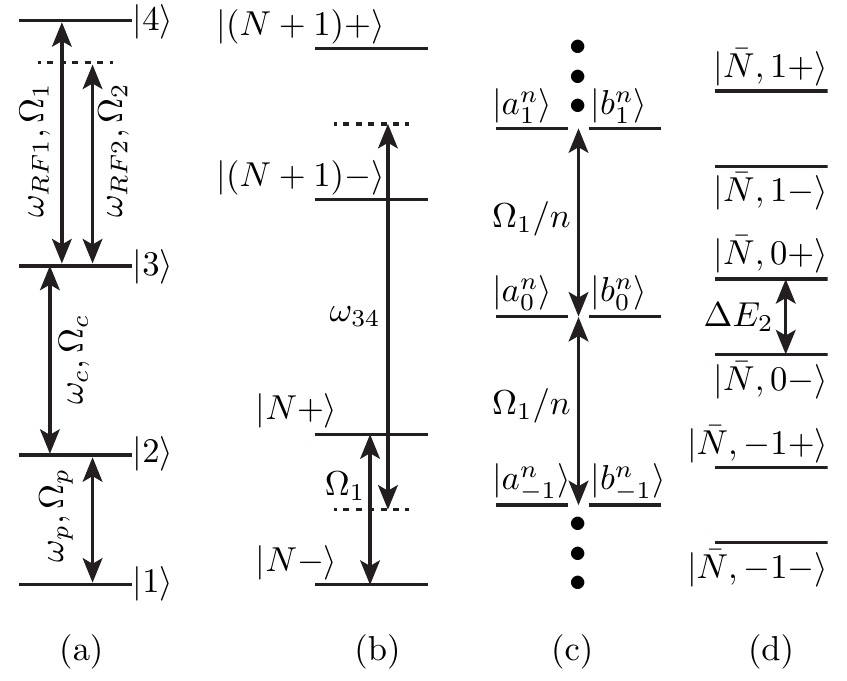}%
    \vspace*{-3mm} 
    \caption{(a) 4-Level system with bichromatic RF fields. Energy level diagram of the RF dressed Rydberg states. (b) Singly dressed system. (c) Double dressing, for $n=-1$, the uncoupled dressed states are degenerate. (d) The doubly dressed states represented in the coupled basis with combined photon number, $\bar{N} = N+M$. The coupling to RF2 splits the degenerate states by energy, $\Delta E_2$. }
    \label{fig:4levelscheme}
\end{figure}

The 4-level EIT scheme is shown in the Fig.
~\ref{fig:4levelscheme}(a). The probe (p) and the coupling (c) beams are configured for Rydberg EIT. The energy spacing between the two Rydberg states, $\ket{3}$ and $\ket{4}$, is in the RF range. We realize a bichromatic field by addressing the two Rydberg levels with two near resonant RF fields. One of the RF fields, the in-band field, RF1, is kept on resonance while the other field, the out-band field, RF2, is kept a few tens of MHz away from the resonance.
The total Hamiltonian that governs the 4-level system is described as $H = H_A + H_{RF1} + H_{RF2}$, where $H_{A}$ denotes the bare atom Hamiltonian interacting with the probe and the coupling beams. $H_{RF1}$ and $H_{RF2}$ describe the two RF fields interacting with the atom. The coupling strengths for the transitions are $\Omega_p$, $\Omega_c$, $\Omega_1$ and $\Omega_2$ and the detunings are $\Delta_p = \omega_{12}-\omega_p$, $\Delta_c = \omega_{23}-\omega_c$ and $\Delta_{1(2)} = \omega_{34}-\omega_{RF1(2)}$. The full Hamiltonian in matrix form is shown in Appendix~\ref{app:Hamiltonian}.

The dynamics of the system are calculated using Liouville's equation, $\dot{\rho} = i / \hbar [\rho,H] + \mathcal{L}(\rho)$, where $\mathcal{L} = \mathcal{L}_{decay} + \mathcal{L}_{tt} +\mathcal{L}_{laser}$ is the Lindblad operator. The three terms account for the atomic decay rate, the transit time broadening and the laser dephasing rate, respectively. A steady state solution is obtained by numerically solving the Liouville's equation. We average over the thermal velocity distribution to match our room temperature experiment.

The experimental setup consists of a Cs vapor cell with counter-propagating probe and coupling laser beams. We tune the probe beam to the D2 transition, $\lambda_p = 852$\,nm. The coupling laser is tuned to address the $56D_{5/2}$ at $\lambda_c=509$\,nm. The RF fields, addressing the $56D_{5/2}$ to $57P_{3/2}$ transition of Cs, are incident on the vapor cell from an orthogonal direction. RF1 and RF2 are emitted from the same antenna in order to maximize the field overlap. The polarization of the two laser beams and the RF sources are linear and aligned in parallel. The D to P transition can be split without any residual uncoupled transition probability in the presence of RF1 alone~\cite{Sedlacek:2013}. Two probe transmission peaks are observed when the coupling laser is tuned across the resonance.

\begin{figure}
    \centering
    \includegraphics[width=\linewidth]{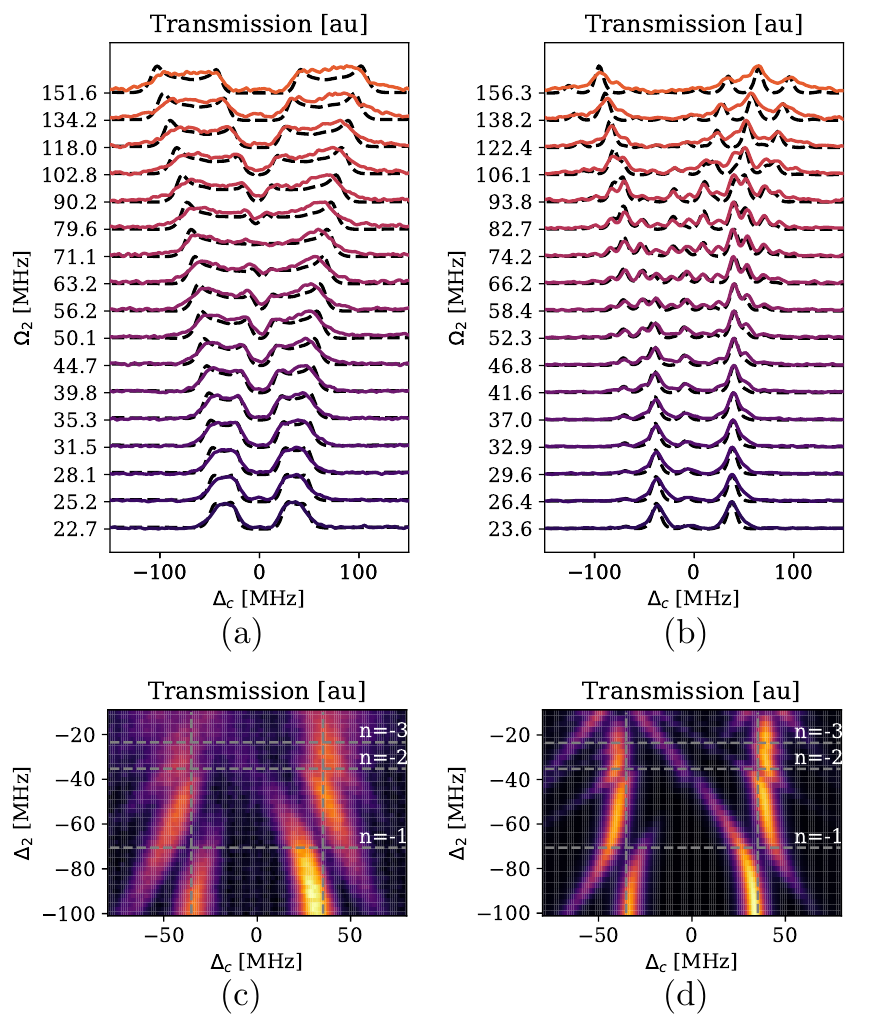}
    \vspace*{-8mm} 
    \caption{Transmission data (solid lines) for $\Omega_2$ scan and calculations (dashed lines). (a) $\Delta_2  = -1$\,MHz and (b) $\Delta_2 = -30$\,MHz. (c) Experiment and (d) calculation for $\Delta_2$ scan at $\Omega_1 = 74.5$\,MHz. Multiple avoided crossings occur as $\Delta_2$ matches the harmonic and sub-harmonic conditions. The horizontal dashed lines mark the values of n, $\Delta_{2} = \Omega_{1} / n$. Vertical dashed lines mark the singly dressed state Rabi splitting, $\pm\Omega_{1}/2$.}
    \label{fig:Out_band_RF_power}
\end{figure}

The results of the calculation show good agreement with the data, Fig.~\ref{fig:Out_band_RF_power}. We vary the out-band power, $\Omega_2$, for different out-band detunings, $\Delta_2$. Small deviations of the data relative to the calculations can arise from inhomogeneous and stray electric DC and RF fields.

The spectra are best understood using a dressed state picture where the two Rydberg states, $\ket{3}$ and $\ket{4}$, are dressed by the RF fields. We first dress the Rydberg states with RF1, which is on resonance, i.e., $\omega_{RF1}=\omega_{34}$. The RF1 dressed states are then dressed by RF2. The doubly dressed picture reveals the conditions for harmonic and sub-harmonic resonances~\cite{Agarwal:91, Rudolph:98}.

The singly dressed states are described by uncoupled atomic and photonic states, 
\begin{equation}
    \ket{N\pm} = \frac{1}{\sqrt{2}}(\ket{3,N} \pm \ket{4, N-1}),
\end{equation}
where $\ket{r,N}$ is comprised of the Rydberg states $\ket{r}$ (r=3,4) and photon number state $\ket{N}$, with $N$ the number of excitations present in the RF1 field mode. The Hamiltonian of the Rydberg states interacting with RF1 is $H_1 = H_0 + H_{RF1}$, where
\begin{equation}
    H_1 \ket{N\pm} = \hbar (N\omega_{34} \pm \Omega_1/2)\ket{N\pm}.
\end{equation}
The dressed states, $\ket{N+}$ and $\ket{N-}$  form a ladder of doublets separated by $\omega_{34}$ with intra-doublet spacing $\Omega_1$, Fig.~\ref{fig:4levelscheme}(b).

The addition of RF2 modifies the RF1 Jaynes-Cummings ladder. For $M$ excitations in RF2, the uncoupled doubly dressed states are $\ket{N \pm, M}$ with energy, $E\pm = \hbar (N\omega_{34}\pm \Omega_1/2 + M\omega_{RF2})$. Exchange of RF2 excitations, $m$ and $m'$ with the singly dressed system, $\ket{N +}$ and $\ket{N -}$, results in states,
\begin{equation}
    \begin{aligned}
    \ket{+} = &\ket{(N-m')+, M+m'},\\
    \ket{-} = &\ket{(N-m)-, M+m},
    \end{aligned}
\end{equation}
where $m'$ or $m$ are the excitation number loss (gain) from the singly dressed states and gain (loss) in the RF2 mode. The energies are
\begin{equation}
    \begin{aligned}
    E_+ &= \hbar [ (N-m')\omega_{34} + \Omega_1/2 + (M+m')\omega_{RF2} ], \\
    E_- &= \hbar [ (N-m)\omega_{34} - \Omega_1/2 + (M+m)\omega_{RF2}] .
    \end{aligned}
\end{equation}

We focus on the regime where the two states are degenerate, i.e., $E_+ = E_-$. In this case, the states cross when $\Delta_{2} = \Omega_1/n$, where $n = m-m'$ is a non-zero integer. The expression indicates that for fixed $n$, $\ket{+}$ and $\ket{-}$ cross as a function of $\Delta_{2}$ and $\Omega_1$. It is convenient to redefine the states by substituting $m' = m - n$ to form the basis
\begin{equation}
\begin{aligned}
    \ket{(N+n-m)+, M-n+m} & \equiv \ket{a_m ^n}, \\   
    \ket{(N-m)-, M+m} & \equiv \ket{b_m ^n}.
\end{aligned}
\end{equation}
The energy spacing between the ladder of these degenerate states for two adjacent $m$ values is $\Omega_1/n$, Fig.\ref{fig:4levelscheme}(c).
 
The coupling between RF2 and the singly dressed states is captured by treating the interaction perturbatively with perturbation parameter, $\lambda = \Omega_2 / \Omega_1$. For $|n| = 1$ the solution is straight forward. For other $|n|$, the couplings are more complicated but perturbative expressions can be derived. The interaction causes avoided energy level crossings with energy splitting,
\begin{equation}\label{eq:double_splitting}
    \Delta E_2 = \hbar \frac{\Omega_1}{2} 
    \Big[ \Big(\frac{2\delta}{\Omega_1} - \frac{1}{8}\Big(\frac{\Omega_2}{\Omega_1}\Big)^2 \Big)^2 + \Big(\frac{\Omega_2}{\Omega_1}\Big)^2 \Big]^\frac{1}{2},
\end{equation}
where $\delta = \Delta_2-\Omega_1/n$ is the detuning of RF2 from the harmonic resonance~\cite{Rudolph:98}. Higher orders of perturbation theory can be used to extend the range of validity with respect to the perturbation parameter. The coupled, doubly dressed states are shown in Fig.~\ref{fig:4levelscheme}(d) with energy splitting $\Delta E_2$. Eq.~\ref{eq:double_splitting} has a minimum splitting of $\Omega_2$ when $\delta = \delta_{min} =\Omega_2^2/(16\Omega_1)$. Eq.~\ref{eq:double_splitting} demonstrates that for a detuning near the harmonic resonance, i.e., $\Delta_{2} \sim \Omega_1/n$, the energy splitting of the two doublets changes with effective Rabi frequency, $\Omega_2' = \sqrt{\delta'^2 + \Omega_2^2/4}$, where $\delta' = \delta - \delta_{min}$.

For atom-based RF field sensing, the most important conclusion is that the minimum splitting of the doubly dressed states is invariant to the in-band RF1 field and only depends on the perturbative field resonant Rabi frequency. The intra-doublet splitting at the resonance is $\Delta E_2^{res} = \hbar \Omega_2/2$. The doubly dressed system provides a calibration-free method to measure the power of the perturbative field. Furthermore, RF1 offers a knob to tune the resonance frequency for sensing the perturbative field, RF2. The shape of the avoided crossing can be used to determine SI traceable properties of RF2 and RF1. It is possible to use other sub-harmonic resonances and tune other parameters, like $\Delta_c$, to determine the properties of the RF fields.


When RF2 is applied, a set of different transmission peaks emerge as a function of $\Delta_c$. At $\Delta_2 \lesssim \Gamma$, where $\Gamma$ is the EIT linewidth, the bichromatic EIT signal is broadened as the RF2 power increases, Fig.~\ref{fig:Out_band_RF_power}(a). The sum of the two fields produces an effective field which oscillates at the average frequency and has a slowly beating amplitude. For small detuning, the beat rate can achieve steady state. The total amplitude swings between the sum and difference of the two field amplitudes. When the two powers are equal, the minimum splitting reaches 0 MHz, as can be seen in Fig.~\ref{fig:Out_band_RF_power}(a) at $\Omega_2 \sim 74.5$\,MHz. At larger RF2 detuning the dynamics are different. The atoms see two different frequency components at any given time. The full picture requires a complete Hamiltonian with two RF components, Fig.~\ref{fig:Out_band_RF_power}(b).
In the case of larger RF2 detunings, there is an interplay between the RF1 Rabi frequency and the RF2 detuning. Sub-harmonic resonances occur when, $\Delta_{2} = \Omega_1/n$. Figures~\ref{fig:Out_band_RF_power}(c,d) show a comparison between the experimental results and calculations. The avoided crossings occur at the resonance condition, i.e., for $n=-1,-2,-3$. The slight mismatch between the data and calculations can be attributed to broadening and shifts due to RF field inhomogeneity and stray fields.

\begin{figure}
    \includegraphics[width=\linewidth]{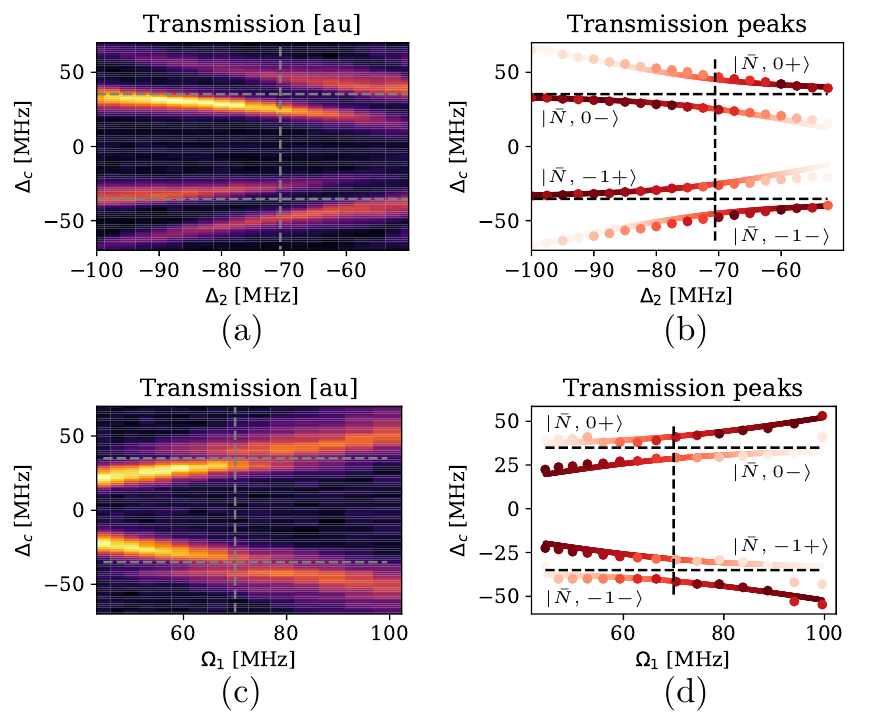}
    \vspace*{-8mm} 
    \caption{Contour map of spectra as a function of (a) $\Delta_2$ at $\Omega_2=19.8$\,MHz, $\Omega_1=70.6$\,MHz and (c) $\Omega_1$ at $\Delta_2 = -70.0$\,MHz, $\Omega_2=12.5$\,MHz. The peak positions are plotted with the amplitude in the color depth, (b) and (d). The circles are peak values obtained from data. The lines overlay the avoided crossing model considering only the first harmonic resonance.}  
    \label{fig:Avoided_crossing_OB_detuning}
\end{figure}

Detailed analysis is performed on the avoided crossing features near the 1st harmonic, i.e. $n=-1$. Two independent experiments are performed: (1) $\Delta_2$ is scanned near $\Omega_1/n$, Fig.~\ref{fig:Avoided_crossing_OB_detuning}(a,b) and (2) $\Omega_1$ is scanned near $\Delta_2$, Fig.~\ref{fig:Avoided_crossing_OB_detuning}(c,d). The two cases represent the scanning of the detuning, $\delta$ across $0$\,MHz in Eq.~\ref{eq:double_splitting} producing harmonic resonance conditions. The transmission peaks display avoided crossing features. The minimum splitting between the transmission peaks is equal to $\Omega_{2}/2$ as if RF2 is on resonance. The condition preserves SI-traceability~\cite{Rudolph:98}. Additionally, the two cases estimate $\Omega_1$ and $\Delta_2$, respectively. Additional examples are shown in Appendix~\ref{app:delta2scan} and \ref{app:omega1scan}.

\begin{figure}
    \centering
    \includegraphics[width=1.0\linewidth]{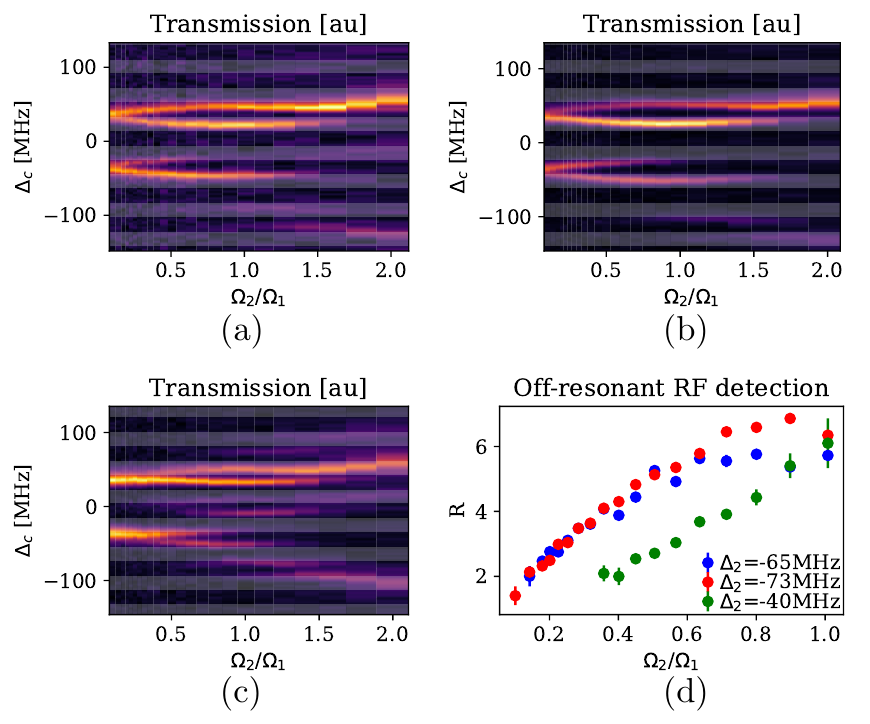}
    \vspace*{-8mm}
    \caption{$\Omega_2$ scan at (a) $\Delta_2=-65$\,MHz, (b) $\Delta_2=-73$\,MHz and (c) $\Delta_2=-40$\,MHz with $\Omega_1 = 72.9$\,MHz. Plot (d) shows average resolution, R of each doublet pairs. $\Delta_2 = 73$\,MHz produces best detectable peaks which corresponds to an enhanced sensitivity of $\sim12$\,dB over the non-resonant condition. The RF1 field is $\sim 11$\,mV/cm whereas the lowest detected RF2 field is $\sim 1.1$\,mV/cm. The double peak near $\Delta_c = 0$\,MHz (c) emerges as $\Delta_2 \sim \Omega_1/n$ with $n=-2$. }
    \label{fig:Out_band_RF_sensitivity}
\end{figure}

The sub-harmonic resonances can be utilized to obtain improved amplitude sensitivity of an off-resonant field by putting the dressed state energy level in resonance. We observe an enhanced sensitivity for $\Omega_1$, $\Omega_2$ and $\Delta_2$ when $\Delta_2$ matches the harmonic resonance conditions. The RF2 detuning is set to -73\,MHz, which is close to the RF1 Rabi frequency. RF power as low as -20\,dBm can be detected whereas in the off resonant case we are limited to -8\,dBm, Fig.~\ref{fig:Out_band_RF_sensitivity}. The spectral resolution of the pair of the peaks split by RF2 is plotted in Fig.~\ref{fig:Out_band_RF_sensitivity}(d). The resolution, $R = 2 |\bar{x}_{1} - \bar{x}_{2}| / (\sigma_{1} + \sigma_{2})$, where $\bar{x}_i$ and $\sigma_i$ with $i\in[1,2]$ are the mean and standard deviation is obtained by fitting a double Gaussian to each pair of split peaks. At $R=1$, the two peaks are separated by $\sigma_1 + \sigma_2$. For $R\geq 1$ the peaks are resolved.  

The appearance of equispaced doublets is a trademark of the double Rydberg state dressing. Figures~\ref{fig:Out_band_RF_sensitivity}(a,b) show the condition $n \sim -1$.  The emergence of a doublet in the middle of the main AT peaks at $\Delta_c = 0$\,MHz fulfills the condition $n \sim -2$, Fig.~\ref{fig:Out_band_RF_sensitivity}(c). For $\Omega_2 < \Omega_1$, the doublet splitting increases with $\Omega_2$. When $\Omega_2/\Omega_1 > 1$ the doublet splitting starts to decrease. The effect results from the role reversal of the dressing fields. As $\Omega_2/\Omega_1 > 1$, it is convenient to think of the system as first dressed by RF2 and then by RF1. In the absence of RF1, the AT peaks would be asymmetric with peaks at $\Delta_c \sim 50$\,MHz and $\Delta_c \sim -100$\,MHz with AT splitting of $\sim$\,150MHz, Fig.~\ref{fig:Out_band_RF_sensitivity}(a,b). RF1 is shifted by $=+73$\,MHz from RF2 and fulfils the sub-harmonic condition with $n \sim +2$. Thus, the doublet peak near $\Delta_c \sim -50$\,MHz emerges. The same can be observed in plot~\ref{fig:Out_band_RF_sensitivity}(c) where  $\Delta_1$ is shifted by $40$\,MHz from RF2. Near $\Omega_c \sim 100$\,MHz, the sub-harmonic condition is fulfilled with $n\sim+3$, thus two doublet peaks emerge between the main AT peak splitting. When RF1 and RF2 have equal power, the combined field oscillates as $\sin((\omega_{34} - \Omega_2) t)$. The average frequency is shifted to the center of the two bands, thus the observed AT doublets show asymmetric peak height. Appendix~\ref{app:omega2scan} shows several other examples.

In this work, we demonstrated a doubly dressed Jaynes-Cummings picture of Rydberg states coupled to two near-resonant RF fields. Our study explores the interaction of multiple RF fields in Rydberg atom sensors beyond the mixer model. We show that the exchange of excitations between the two fields plays an important role in revealing the spectral features. The double dressing picture is utilized to recover SI traceable RF-based sensing. For sensing applications, enhancement in sensitivity to a non-resonant RF field is demonstrated. The model is utilized as an approach for measuring multiple RF fields by means of mapping avoided crossings. A modulated RF field contains multiple frequency components. Our study paves the way to obtain information associated with each frequency component of a modulated signal. During the final preparation of this manuscript a related paper appeared on the archive~\cite{Jayaseelan:2023}. In that work, a similar problem is addressed, but the paper does not describe the analytic formulas derived from the Jaynes-Cummings analysis of the avoided crossings that enables SI-traceability. The connection between the avoided crossings occurring in the multichromatic RF field and SI-traceability is the main point of our work.


\clearpage
\appendix

\setcounter{equation}{0}
\setcounter{figure}{0}
\setcounter{table}{0}
\setcounter{page}{1}
\makeatletter
\renewcommand{\theequation}{A\arabic{equation}}
\renewcommand{\thefigure}{A\arabic{figure}}

\section{Total Hamiltonian}
\label{app:Hamiltonian}
The total Hamiltonian that governs the 4-level system interacting with probe, coupling, RF1 and RF2 fields is described as: 
\begin{equation}
    H = \hbar \begin{bmatrix}
    0 & \frac{\Omega_p}{2} & 0 & 0\\
    \frac{\Omega_p}{2} & -\Delta_p & \frac{\Omega_c}{2} & 0\\
    0 & \frac{\Omega_c}{2}  & -\Delta_p-\Delta_c  & \frac{\Omega_{1} + e^{-iS(t)}\Omega_{2}}{2} \\
    0 & 0 & \frac{\Omega_{1} + e^{+iS(t)}\Omega_{2}}{2} & -\Delta_p-\Delta_c-\Delta_{1}
    \end{bmatrix},
    \label{eq:H_total}
\end{equation}
where $\Delta_p = \omega_{12}-\omega_p$, $\Delta_c = \omega_{23}-\omega_c$ and $\Delta_1 = \omega_{34}-\omega_{RF1}$ are the detunings of the probe, coupling and RF1 fields from the corresponding transition frequencies, $\omega_{ij}$ ($i,j \in [1,2,3,4]$). As we arbitrarily chose relative phase = 0, the time dependent phase factor, $S(t)= 2 \pi \Delta_2 t$ where $\Delta_2 = \omega_{34}-\omega_{RF2}$ is the detuning of the RF2 field. The coupling strengths for the transitions are the Rabi frequencies, i.e., $\Omega_p$, $\Omega_c$, $\Omega_1$ and $\Omega_2$.

\section{$\Delta_2$ scan avoided crossing}
\label{app:delta2scan}
\begin{figure}[ht]
    \centering
    \includegraphics[width=1.0\linewidth]{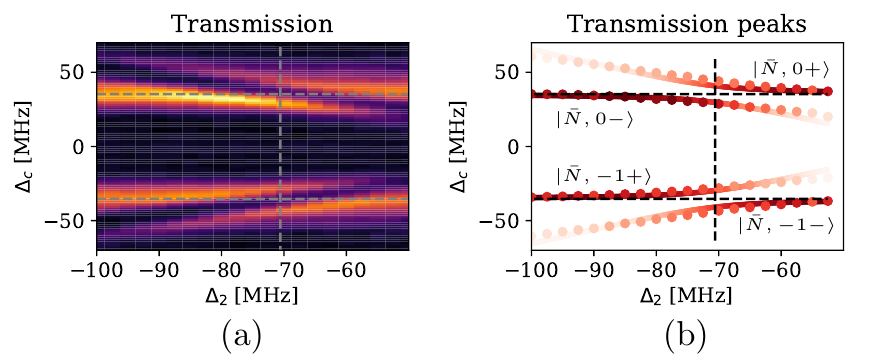}
    \vspace*{-6mm}
    \caption{(a) Experimental data of a transmission map generated by scanning $\Delta_2$ and $\Delta_c$ at $\Omega_2 = 11.2$ and $\Omega_1=70.6$\,MHz. (b) Avoided crossing model (solid line) is plotted against the peak positions obtained from the data (circles).}
    \label{fig:double_dressed_states_delta2_scan}
\end{figure}

In Fig.~\ref{fig:double_dressed_states_delta2_scan}, $\Delta_2$ is scanned while keeping $\Omega_2 = 11.2$\,MHz and $\Omega_1 = 70.6$\,MHz. Peak positions are obtained by fitting a double Gaussian model. Figure~\ref{fig:double_dressed_states_delta2_scan} is an additional example, as per main text, Fig.~\ref{fig:Avoided_crossing_OB_detuning}(a,b). These measurements demonstrate a way to estimate $\Omega_1$ with the knowledge of $\Delta_2$ by observing either of the avoided crossings in cases, where scan range of $\Delta_c$ is limited due to laser performance.

\section{$\Omega_1$ scan avoided crossing}
\label{app:omega1scan}
\begin{figure}[ht]
    \centering
    \includegraphics[width=1.0\linewidth]{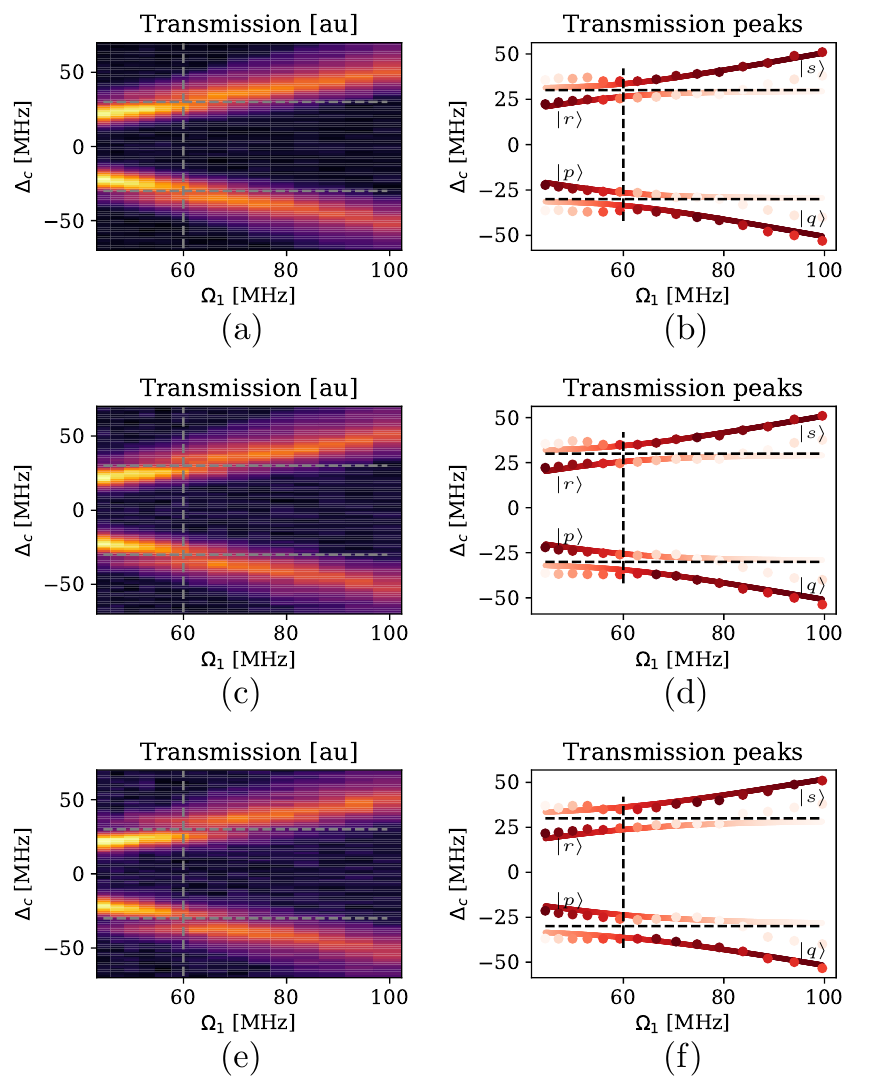}
    \vspace*{-6mm}
    \caption{Contour map of $\Omega_1$ scan near $\Delta_2 = -60$\,MHz with $n=-1$ at (a) $\Omega_2 = 7.1$\,MHz, (c) $\Omega_2 = 8.9$\,MHz and (e) $\Omega_2 = 12.5$\,MHz. The corresponding peak positions (b), (d), and (f) are plotted with the amplitude shown in the color depth for the same parameters as the contour maps. The circles are peak values obtained from the experimental data whereas the solid lines are avoided crossing energy levels. The doubly dressed states are $\ket{p} = \ket{\bar{N}, -1+} $, $\ket{q} = \ket{\bar{N}, -1-} $, $\ket{r} = \ket{\bar{N}, 0-} $ and $\ket{s} = \ket{\bar{N}, 0+} $.}
    \label{fig:double_dressed_states_omega1_scan}
\end{figure}

In Fig.~\ref{fig:double_dressed_states_omega1_scan}, $\Omega_1$ scanned while keeping $\Delta_2 = -60$\,MHz. Peak positions are obtained by fitting a double Gaussian model. Figure~\ref{fig:double_dressed_states_omega1_scan} shows additional data in support of Fig.~\ref{fig:Avoided_crossing_OB_detuning}(c,d) of the main text. The appearance of the avoided crossings at a significantly small $\Omega_2$ ($\sim 0.1 \Omega_1$) proves this technique useful for weak off-resonant RF field sensing. Here, RF1 acts as a control field and it tunes the atomic (dressed) states to match $\Delta_2$. Due to this resonance condition the RF field sensing gets enhanced for RF2.

\section{$\Omega_2$ scan avoided crossing}
\label{app:omega2scan}
\begin{figure}[ht]
    \centering
    \includegraphics[width=1.0\linewidth]{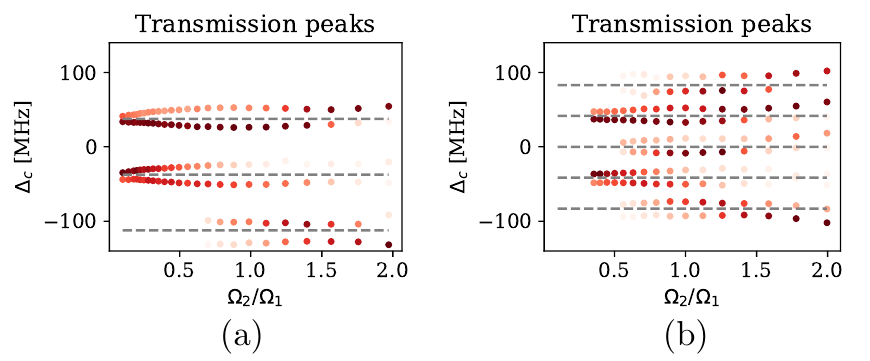}
    \vspace*{-6mm}
    \caption{Peak positions for double dressed states. $\Omega_1 = 72.9$\,MHz (a) $\Delta_2=-73$\,MHz and (b) $\Delta_2 = -40$\,MHz.}
    \label{fig:double_dressed_states_omega2_scan}
\end{figure}

In Fig.~\ref{fig:double_dressed_states_omega2_scan}, $\Omega_2$ is scanned while keeping $\Omega_1$ fixed at $72.9$\,MHz. Peak positions are obtained by fitting a double Gaussian model, Fig~\ref{fig:Out_band_RF_sensitivity}(b,c). The equispaced doublets show the energies of the doubly dressed states. Doublet spacing increases for $\Omega_2 < \Omega_1$. The role of the two fields gets exchanged when $\Omega_2 > \Omega_1$. RF2 can be considered to asymmetrically dress the atomic states first which gets doubly dressed by RF1. As a result, the doublet spacing can be seen to increase as $\Omega_2$ decreases. 

\pagebreak

\bibliography{bibfile}

\end{document}